\documentclass[hyper]{JHEP} 

\usepackage{epsfig}

\newcommand\fverb{\setbox\pippobox=\hbox\bgroup\verb}

\newcommand\fverbdo{\egroup\medskip\noindent%

            \fbox{\unhbox\pippobox}\ }

\newcommand\fverbit{\egroup\item[\fbox{\unhbox\pippobox}]}

\newbox\pippobox

\title{Covariant Hamiltonian Formalism for $F(R)-$Gravity}

\author{J. Kluso\v{n}\\
	Department of
	Theoretical Physics and Astrophysics\\
	Faculty of Science, Masaryk University\\
	Kotl\'{a}\v{r}sk\'{a} 2, 611 37, Brno\\
	Czech Republic\\
	E-mail: \email{klu@physics.muni.cz}}

\author{
	B. Matou\v{s} \\
	Department of
	Theoretical Physics and Astrophysics\\
	Faculty of Science, Masaryk University\\
	Kotl\'{a}\v{r}sk\'{a} 2, 611 37, Brno\\
	Czech Republic\\
	E-mail: \email{426945@mail.muni.cz}}

 \abstract{In this short note we perform covariant Hamiltonian analysis
 of $F(R)-$gravity.}

\def\be{\begin{equation}}

\def\ee{\end{equation}}

\def\bea{\begin{eqnarray}}

\def\bN{\mathbf{N}}
\def\eea{\end{eqnarray}}

\def\mH{\mathcal{H}}

\newcommand{\mL}{\mathcal{L}}

\begin{document}
\section{Introduction and Summary}\label{first}
The fundamental concept of modern field theories is their formulation with the help of
action and corresponding Lagrangian density. This formulation is manifestly covariant 
which reflects basic concept of general relativity. Dynamics of these fields is then governed by Lagrange equations of motion which are again manifestly covariant. In other words, 
in the Lagrangian formulation all space-time coordinates are treated on equal footing.

The situation is different when we switch to the Hamiltonian formalism when one direction known as time  direction is selected and Hamiltonian equations of motion  basically determine evolution of the system along this direction. Clearly such a splitting of space-time into space and time
breaks manifest covariance of the theory.
 On the other hand there exists covariant Hamiltonian formulation of field theory known as Weyl-De Donder theory
\cite{DeDonder,Weyl}. Let us demonstrate its principle on the simple example of the Lagrangian for scalar field $\phi$ that clearly depend on  derivative $\partial_a\phi$. In conventional canonical formalism we consider time derivative as special one and define conjugate momentum as derivative of Lagrangian density with respect to $\partial_t\phi$. In case of Weyl-DeDonder theory we treat all 
 partial derivatives on the equal footing which clearly preserves diffeomorphism invariance. Then covariant canonical Hamiltonian
density depends on conjugate momenta $p^a_M$ which are variables conjugate to $\partial_a \phi^M$.  This approach is also known as
multisymplectic field theory, see for example
\cite{Struckmeier:2008zz,Kanatchikov:1997wp,Forger:2002ak}, for review, see \cite{Kastrup:1982qq} and for recent interesting application of this formalism in string theory, see \cite{Lindstrom:2020szt,Kluson:2020pmi}. 

Intuitively it is clear that such a covariant Hamiltonian approach could be very convenient for all covariant theories and especially it would be very appropriate in case of general relativity. In fact, covariant Hamiltonian formulation of gravity was presented by P. Horava in his paper \cite{Horava:1990ba}. It was shown there that the analysis simplifies considerably when instead of conventional canonical variable $g_{ab}$ we use  $f^{ab}\equiv \sqrt{-g}g^{ab}$. In fact, an importance of this variable in the holographic formulation of gravity was stressed recently by T. Padmanabhan in his works, for very clear and detailed discussion of properties of gravity formulated in $f^{ab}$ variables we recommend his paper \cite{Parattu:2013gwa}.

It is natural to ask the question whether covariant Hamiltonian analysis can be performed with more general theories of gravity, as for example $F(R)-$gravity, for review and extensive list of references see \cite{Nojiri:2017ncd,Nojiri:2010wj,DeFelice:2010aj}. These theories can be considered
as simplest modification of general relativity that have a potential to explain some aspects of inflation or dark energy. On the other hand considering $F(R)-$gravity  as generalization of general relativity, it is natural to study whether it is possible to find its covariant Hamiltonian formulation. In fact, this is the goal of this paper.  As the first step we 
 introduce auxiliary fields in order to be able to deal with $F(R)-$gravity and then we carefully split Lagrangian density for $F(R)-$gravity into bulk and boundary part. We find that there is an additional contribution to the bulk part with respect to ordinary gravity that
is proportional to the partial derivative of the auxiliary field. Then we 
can straightforwardly proceed to the covariant Hamiltonian formulation when we firstly determine momenta conjugate to $g_{ab}$. We find that they are rather complicated which has been  previously stressed in \cite{Parattu:2013gwa}. Then, following discussion presented in that paper, we proceed to the canonical variable $f^{ab}=\sqrt{-g}g^{ab}$ that are related to $g_{ab}$ through point transformations. We determine momenta $N^c_{ab}$ conjugate to $f^{ab}$ and $p^c_B$ conjugate to auxiliary field $B$. Finally we invert relations between conjugate momenta $N^c_{ab}$ and Christoffel symbols and also between $p^c_B$ and derivative of $B$ and obtain corresponding Hamiltonian. We also determine canonical equations of motion. 

Let us outline our results. We find covariant Hamiltonian for $F(R)-$gravity which is the first step in the application of this formalism to more general form of gravity. We also determine corresponding equations of motion.
 It would be certainly nice to extend this result along directions that are related to Padmanabhan's work. It would be also nice to apply covariant Hamiltonian analysis in case of more general theories of gravity, as for example curvature-squared gravity 
\cite{Kluson:2013hza}. Finally, it would be interesting to find
covariant Hamiltonian formalism for $F(R)-$gravity formulated
in Jordan frame and compare it with the result derived in this paper  and perform similar
analysis as in 
\cite{Bahamonde:2016wmz,Bahamonde:2017kbs,Nashed:2019yto}.
 We hope to return to these problems in future.

This paper is organized as follows. In the next section (\ref{second}) we introduce Lagrangian for $F(R)-$gravity and split it into bulk and boundary parts. Then in section 
(\ref{third}) we present canonical formalism in $g_{ab}$ and $M^{abc}$ variables. Finally in section (\ref{fourth}) we proceed to the Hamiltonian analysis in $f^{ab},N^c_{ab}$ variables and determine corresponding Hamiltonian density and canonical equations of motion.

\section{Introduction of $F(R)$-Gravity}\label{second}
In this section we introduce an action for $F(R)-$gravity. This theory is generalization 
of Einstein-Hilbert action when term linear in curvature $R$ is replaced by arbitrary function 
$F(R)$. Explicitly, the action for $F(R)-$gravity has the form
\begin{equation}\label{actF}
S=\int d^4x \mL^F \ ,  \quad \mL^F=\frac{1}{16\pi}\sqrt{-g}F(R) \ , 
\end{equation}
where we set gravitational constant and speed of light  to be equal one and where $F$ is an arbitrary function 
of Ricci scalar $R$ that is defined as
\footnote{We work with the metric $g_{ab}$  with signature $(-1,1,1,1)$ where $a,b,c,\dots=0,\dots,3.$ }
\begin{eqnarray}
& &R=g^{mn}R_{mn}=g^{mn}R_{m k n}^{ \ k} \ ,  \quad  \nonumber \\
& & R^k_{ \ lmn}=\partial_m \Gamma_{n l}^k-\partial_n \Gamma_{m l}^k+\Gamma^k_{mp}\Gamma^p_{n l}-
\Gamma^k_{np}\Gamma^p_{m l} \ . \nonumber \\
\end{eqnarray}
In order to find Hamiltonian formulation of this theory it is convenient
to introduce two scalar fields $A$ and $B$ 
and write the Lagrangian density in the form
\begin{equation}\label{extL}
\mL=\frac{1}{16\pi}\sqrt{-g}[F(B)+A(R-B)] \ , 
\end{equation}
where the equations of motion for $A$ implies $R=B$ while equation of motion for $B$ implies
$F'(B)-A=0$. Inserting the first result into (\ref{extL}) we obtain original action. Instead it is useful to use second equation to express $A=F'(B)$ and write the Lagrangian density (\ref{extL}) in the form 
\begin{equation}\label{actionB}
\mL=\frac{1}{16\pi}\sqrt{-g}[F(B)+F'(B)(R-B)] \ . 
\end{equation}
In order to check consistency of this action let us solve equations of motion for $B$ that
follow from (\ref{actionB}) 
\begin{equation}
F'(B)+F''(B)(R-B)-F'(B)=F''(B)(R-B)=0
\end{equation}
which again implies $R=B$ on condition $F''(B)\neq 0$. Inserting this result into (\ref{actionB}) we obtain original Lagrangian density (\ref{actF}). In fact, for $F''(B)=0$ we find that $F'(B)=K$ where $K=\mathrm{const}$ and hence $F(B)=KB+C$ where $C$ is another constant. Inserting this result into (\ref{actionB}) we obtain Einstein-Hilbert action with cosmological constant proportional to $C$ after appropriate rescaling. In what follows we will presume condition $F''(B)\neq 0$ that leads to $F(R)-$gravity.

Following \cite{Parattu:2013gwa} we write  Ricci scalar as 
\begin{equation}
R=
Q_k^{ \ mnl}R^k_{\ \ mnl} \ , 
\end{equation}
where
\begin{equation}
Q_k^{ \ mnl}=\frac{1}{2}(g^{ml}\delta_k^n-
g^{mn}\delta_k^{l}) \ , \quad  Q_k^{ \ mnl}=-Q_k^{\ mln} \ .
\end{equation}
As the next step we split Lagrangian density (\ref{actionB})
into bulk and boundary part \cite{Eddington,Schrodinger}, for recent careful discussion, 
see also \cite{Parattu:2013gwa}\footnote{It is important to stress that the variation of the bulk term alone leads to the equations of motion  while the surface term, when integrated over a horizon, is
related to the entropy of the horizon, as was shown for example in 
\cite{Padmanabhan}.}.
 This procedure is trivial in case of terms quadratic in Christoffel symbols in Riemann tensor however we should be more careful with terms that contain partial derivatives of Christoffel tensor since for example
\begin{eqnarray}
& &\sqrt{-g}F'(B)Q_k^{ \ mnl}\partial_m \Gamma^k_{nl}=
\partial_m[\sqrt{-g}F'(B)Q_k^{ \ mnl}\Gamma^k_{nl}]-
\nonumber \\
& &-\partial_m[\sqrt{-g}Q_k^{ \ mnl}]F'(B)\Gamma^k_{nl}-
\sqrt{-g}Q_k^{ \ mnl}F''(B)\partial_m B \Gamma^k_{nl} \ . 
\end{eqnarray}
The second term on the second line gives bulk contribution corresponding
to the kinetic term for auxiliary field $B$. Further, the first term gives
also bulk contribution as follows from the fact that 
\begin{equation}
\partial_c (\sqrt{-g}Q_a^{ \ bcd})=-\sqrt{-g}\Gamma^b_{cm}Q_a^{ \ mcd}
+\Gamma^m_{ca}Q_m^{ \ bcd}\sqrt{-g} 
\end{equation}
which is consequence of the condition $\nabla_c (\sqrt{-g}Q_a^{ \ bcd})=0$.
%
Collecting these terms together we obtain that surface term has explicit form 
\begin{eqnarray}
& &16\pi\mL_{surf}^F=2\partial_c[\sqrt{-g}F'(B)Q_a^{ \ bcd}\Gamma_{bd}^a]=\nonumber \\
& & =2F''(B)\partial_c B Q_a^{ \ bcd}\Gamma^a_{ \ bd}-4F'(B)
\sqrt{-g}Q_a^{ \ bcd}\Gamma_{dm}^a\Gamma^m_{bc}+2F'(B)
\sqrt{-g}Q_a^{ \ bcd}\partial_c \Gamma^a_{bd}
\nonumber \\
\end{eqnarray}
and hence we find that decomposition of the Lagrangian density of $F(R)-$gravity 
into  surface and bulk terms has the form 
\begin{eqnarray}\label{mLfinal}
& &16\pi\mL^F=
16\pi\mL^F_{quadr}+16\pi\mL^F_{surf} \ , \nonumber \\
& &16\pi\mL^F_{quadr}=
2F'(B)Q_a^{ \ bcd}\Gamma_{dm}^a\Gamma^m_{ \ bc}
-2F''(B)\partial_c B Q_a^{ \ bcd}\Gamma^a_{ \ bcd}
+\sqrt{-g}[F(B)-F'(B)B] \ , \nonumber \\
& &16\pi\mL^F_{surf}=
2\partial_c[\sqrt{-g}F'(B)Q_a^{ \ bcd}\Gamma_{bd}^a] \ . 
\nonumber \\
\end{eqnarray}
The Lagrangian density (\ref{mLfinal}) is our starting point for the
covariant Hamiltonian formulation of $F(R)-$gravity which will be analysed
in the next section.
\section{Covariant Hamiltonian Formalism with $(g_{ab},M^{abc})$ as canonical variables}
\label{third}
In this section we proceed to the covariant Hamiltonian formalism of $F(R)-$gravity when 
we consider $g_{bc}$ as basic dynamical variable. Then, according to the basic principle
of covariant field theory, the conjugate momentum $M^{abc}$ is defined as variation of 
the bulk part of the action with respect to $\partial_a g_{bc}$. Explicitly, we know
that quadratic part of Lagrangian density for $F(R)-$gravity has the form 
(\ref{mLfinal}), or alternatively, using explicit form of  $Q_a^{ \ bcd}$ we have 
\begin{eqnarray}\label{mLexp}
& &16\pi\mL_{quad}^F=\sqrt{-g}F'(B)[\Gamma^h_{dk}\Gamma^k_{gh}g^{gd}-
\Gamma^f_{fk}\Gamma^k_{gh}g^{gh}] -\nonumber \\
& &-F''(B)\sqrt{-g}\partial_cB[\Gamma^c_{bd}g^{bd}-g^{cb}\Gamma_{bd}^d]+\sqrt{-g}[F(B)-F'(B)B] \ . \nonumber \\
\end{eqnarray}
In order to define conjugate momenta $M^{abc}$ we need following expression
\begin{equation}
\frac{\delta \Gamma^k_{bc}}{\delta \partial_p g_{rs}}=
\frac{1}{4}\delta^p_b(g^{kr}\delta_c^s+g^{ks}\delta_c^r)+
\frac{1}{4}\delta^p_c(g^{kr}\delta_b^s+g^{ks}\delta_b^r)-
\frac{1}{4}g^{kp}(\delta_b^r\delta_c^s+\delta_c^r\delta_b^s)\ . 
\end{equation}
Using this result we get  
\begin{eqnarray}\label{Mabc}
 & &16\pi M^{abc}=\frac{\partial \mL_{quad}^F}{\partial ( \partial_a g_{bc})}
=\frac{1}{2}F'(B)\sqrt{-g}[g^{bd}\Gamma_{dk}^ag^{kc}+g^{cd}\Gamma_{dk}^a g^{kb}]
\nonumber \\
& &-\frac{1}{2}F'(B)\sqrt{-g}\Gamma^k_{fk}(g^{fb}g^{ac}+g^{fc}g^{ab})
-\frac{1}{2}F'(B)\sqrt{-g}g^{bc}\Gamma^a_{gh}g^{gh}+\frac{1}{2}F'(B)\sqrt{-g}\Gamma^k_{fk}g^{fa}g^{bc}-
\nonumber \\
& &-\frac{1}{2}F''(B)\sqrt{-g}\partial_f B [g^{fb}g^{ac}+g^{fc}g^{ab}]
+F''(B)\sqrt{-g}\partial_f B g^{fa}g^{bc} \nonumber \\
\end{eqnarray}
and momentum conjugate to $B$ as
\begin{equation}
16\pi p^c_B=\frac{\partial \mL_{quadr}^F}{\partial(\partial_c B)}=
-F''(B)\sqrt{-g}[\Gamma^c_{bd}g^{bd}-g^{cb}\Gamma_{bd}^d] \ . 
\end{equation}
To proceed further let us introduce $V^a\equiv -g_{bc}M^{abc}$ that, using
(\ref{Mabc}) has the form
\begin{eqnarray}
16 \pi V^a
=\sqrt{-g}F'(B)\Gamma^a_{gh}g^{gh}-F'(B)\sqrt{-g}\Gamma^k_{kf}g^{fa}
-3F''(B)\sqrt{-g}\partial_f B g^{fa} \ . \nonumber \\
\end{eqnarray}
From $V^a$ we can express $\partial_f B$ as 
\begin{equation}
\partial_f B=-\frac{16\pi g_{fa}}{3F''(B)\sqrt{-g}}
\left(g_{bc}M^{abc}+\frac{F'(B)}{F''(B)}p^a_B\right) \ . \nonumber \\
\end{equation}
Then we can insert this result into definition of $M^{abc}$ and find relation between
$M^{abc}$ and $\Gamma^a_{bc}$. However resulting expression is rather complicated
and corresponding Hamiltonian as well. For that reason we rather focus on 
set of variables that were used in \cite{Horava:1990ba}.

\section{Hamiltonian Analysis in New Variables $(f^{ab},N^c_{ab})$ }\label{fourth}
We define new metric variable as 
\begin{equation}\label{deff}
f^{ab}=\sqrt{-g}g^{ab} \ . 
\end{equation}
Even if we call it as a new one we should stress that it was
used in classical literature long time ago \cite{Eddington,Schrodinger} and it was
also used by P. Horava for covariant Hamiltonian formulation of gravity
\cite{Horava:1990ba}. An importance of these variables was also stressed many times in 
works by Padmanabhan, see for example \cite{Parattu:2013gwa}. From definition of $f^{ab}$ given in (\ref{deff})
we find that 
\begin{equation}
f\equiv \det f^{ab}=\det g 
\end{equation}
and also we have following relation between variation of $g^{ab}$ and $f^{ab}$
\begin{equation}
\delta g^{ab}=\frac{\delta f^{ab}}{\sqrt{-f}}-\frac{1}{2\sqrt{-f}}f^{ab}
f_{mn}\delta f^{mn}\equiv \frac{1}{\sqrt{-f}}B^{ab}_{mn}\delta f^{mn} \ , 
\end{equation}
where
\begin{equation}
B^{ab}_{mn}=\frac{1}{2}(\delta^a_m\delta^b_n+\delta^b_m\delta^a_n)-
\frac{1}{2}f^{ab}f_{mn}=
\frac{1}{2}(\delta^a_m\delta^b_n+\delta^b_m\delta^a_n-g^{ab}g_{mn}) \ . 
\end{equation}
Note that in our convention $f_{ab}$ is inverse to $f^{ab}$ 
and it has explicit form $f_{ab}=\frac{1}{\sqrt{-g}}g_{ab}$.

Now let us consider $f^{ab}$ as canonical variable and introduce corresponding conjugate
momenta $N^c_{ab}$ as variation of the action with respect to $\partial_cf^{ab}$. Since
boundary term does not contribute to this definition we have
\begin{eqnarray}
N^c_{ \ ab}=\frac{\partial L^F_{quadr}}{\partial( \partial_c f^{ab})}
=\frac{\partial L^F_{quadr}}{\partial(\partial_d g_{mn})}\frac{\partial (
	\partial_d g_{mn})}{\partial(\partial_c f^{ab})} \ . 
\nonumber \\
\end{eqnarray}
Now $f^{ab}$ and $g_{mn}$ are related by point transformations so that 
$f^{ab}=f^{ab}(g_{mn})$ or inverse $g_{mn}=g_{mn}(f^{ab})$. 
Then 
\begin{equation}
\partial_d g_{mn}=\frac{\delta g_{mn}}{\delta f^{ab}}\partial_d f^{ab}
\end{equation}
so that $\partial_d g_{mn}=\partial_d g_{mn}(f^{ab},\partial_c f^{ab})$. Then clearly we have
\begin{eqnarray}
\frac{\delta (\partial_d g_{mn})}{\delta (\partial_c f^{ab})}=
\frac{\delta g_{mn}}{\delta f^{ab}}\delta_d^c \ .  \nonumber \\
\end{eqnarray}
Returning to $N^c_{ \ ab}$ we obtain 
\begin{equation}\label{NM}
N^c_{ \ ab}=\frac{\partial \mL^F_{quadr}}{\partial(\partial_c g_{mn})}(-g_{mk}\frac{1}{\sqrt{-f}}B^{kl}_{ \ ab}g_{ln})=-M^{cmn}\frac{1}{\sqrt{-f}}B_{mn,ab} \ ,  
\end{equation}
where we used 
\begin{equation}
\frac{\delta g_{mn}}{\delta f^{ab}}=
-\frac{1}{\sqrt{-f}}g_{mk}B^{kl}_{\ ab}g_{ln}
\end{equation}
and also we defined $B_{mn,ab}$ as
\begin{equation}
B_{mn,ab}=
g_{mk}B^{kl}_{ab}g_{ln}
=\frac{1}{2}(g_{ma}g_{nb}+g_{mb}g_{na}-
g_{mn}g_{ab}) \ . 
\end{equation}
Since 
\begin{eqnarray}
& &16\pi M^{abc}=\frac{1}{2}F'(B)\sqrt{-g}[g^{bd}\Gamma_{dk}^ag^{kc}+g^{cd}\Gamma_{dk}^a g^{kb}]
\nonumber \\
& &-\frac{1}{2}F'(B)\sqrt{-g}\Gamma^k_{fk}(g^{fb}g^{ac}+g^{fc}g^{ab})
-\frac{1}{2}F'(B)\sqrt{-g}g^{bc}\Gamma^a_{gh}g^{gh}+\frac{1}{2}F'(B)\sqrt{-g}\Gamma^k_{fk}g^{fa}g^{bc}-
\nonumber \\
& &-\frac{1}{2}F''(B)\sqrt{-g}\partial_f B [g^{fb}g^{ac}+g^{fc}g^{ab}]
+F''(B)\sqrt{-g}\partial_f B g^{fa}g^{bc} \nonumber \\
\end{eqnarray}
we obtain from (\ref{NM}) following conjugate momenta $N^c_{ab}$
\begin{eqnarray}\label{Ncab}
& &16\pi N^c_{ab}=
-F'(B)\Gamma^c_{ab}
+\frac{1}{2}F'(B)(\Gamma_{ak}^k\delta^c_b+\Gamma^k_{bk}\delta^c_a)+
\nonumber \\
& &
+\frac{1}{2}F''(B)(\partial_a B \delta_b^c+\partial_b B\delta^c_a)+\frac{1}{2}
F''(B)\partial_g B f^{gc}f_{ab}
\nonumber \\
\end{eqnarray}
and
\begin{equation}\label{pcB}
16 \pi p^c_B=-F''(B)[\Gamma_{bd}^c f^{bd}-f^{cb}\Gamma_{bd}^d]  \ . 
\end{equation}
It is crucial that for $F(B)=B, F'(B)=1 \  , F''(B)=0$ we get that $N^c_{ab}$ has the same
form as in case of pure gravity and that $p^c_B=0$. Further, if we take the trace $f^{ab}N_{ab}^c$ we obtain
\begin{eqnarray}
16\pi f^{ab}N_{ab}^c=-F'(B)\Gamma^c_{ab}f^{ab}+F'(B)\Gamma^k_{kb}f^{bc}+3
F''(B)\partial_g  B f^{gc} \ . 
\end{eqnarray}
In case when $F'(B)=1$ we get trivial identity while for $F''(B)\neq 0$ we can use previous equation and also definition of canonical momenta $p^c_B$ given in 
(\ref{pcB}) to express $\partial_a B$  as 
\begin{eqnarray}
16\pi f^{ab}N_{ab}^c
=16\pi\frac{F'(B)}{F''(B)}p^c_B+3F''(B)\partial_g Bf^{gc}
\nonumber \\
\end{eqnarray}
so that
\begin{equation}
\partial_g B=\frac{16\pi}{3F''(B)}
f_{gc}\left(f^{ab}N_{ab}^c-\frac{F'(B)}{F''(B)}p^c_B\right) \ . 
\end{equation}
Inserting this result into (\ref{Ncab}) we obtain 
\begin{eqnarray}
& &16\pi N_{ab}^c=-F'(B)\Gamma_{ab}^c+\frac{1}{2}F'(B)(\Gamma_{ak}^k \delta_b^c+
\Gamma_{bk}^k\delta_a^c)+\nonumber \\
& &+\frac{16\pi}{6}(f_{ad}f^{mn}N^d_{mn}\delta^c_b+
f_{bd}f^{mn}N_{mn}^d\delta^c_a+f^{mn}N_{mn}^cf_{ab})-\nonumber \\
& &-\frac{16\pi}{6}
\frac{F'(B)}{F''(B)}(f_{ad}p^d_B+f_{bd}p^d_B+p^c_Bf_{ab}) \nonumber \\
\end{eqnarray}
that allows us to write
\begin{equation}\label{bNabc}
16\pi\bN_{ab}^c=-F'(B)\Gamma^c_{ab}+\frac{1}{2}F'(B)
(\Gamma^k_{ak}\delta^c_b+\Gamma^k_{bk}\delta^c_a) \  ,
\nonumber \\
\end{equation}
where
\begin{eqnarray}\label{bN}
& &\bN^c_{ab}=N^c_{ab}
-\frac{1}{6}(f_{ad}f^{mn}N^d_{mn}\delta^c_b+
f_{bd}f^{mn}N_{mn}^d\delta^c_a+f^{mn}N_{mn}^cf_{ab})+\nonumber \\
& &+\frac{1}{6}
\frac{F'(B)}{F''(B)}(f_{ad}p^d_B\delta^c_b+f_{bd}p^d_B\delta^c_a+p^c_Bf_{ab}) \ . 
\nonumber \\
\end{eqnarray}
As the next step we have to find inverse relation between $\Gamma^c_{ab}$ and $\bN^c_{ab}$. 
Following \cite{Parattu:2013gwa} we presume that it has the form
\begin{equation}\label{ans}
\Gamma_{ab}^c=a\bN_{ab}^c+b(\bN_{ad}^d\delta^c_b+\bN_{bd}^d 
\delta_a^c) \ , 
\end{equation}
where $a,b$ are unknown coefficients which we determine after inserting  this ansatz
into (\ref{bNabc}). Explicitly we get 
\begin{eqnarray}
& &\frac{16\pi}{F'(B)}\bN^c_{ab}=
-(a\bN_{ab}^c+b(\bN_{ad}^d\delta^c_b+\bN_{bd}^d
\delta_a^c))
\nonumber \\
& &+\frac{1}{2}[(a
\bN^d_{ad}+b(4\bN^k_{ak}+\bN^k_{ak}))\delta^c_b+
(a\bN^d_{bd}+b(\bN^k_{bk}+4\bN^k_{bk})\delta^c_b)] \ . 
\nonumber \\
\end{eqnarray}
Comparing left and right side of this expression  we obtain following
equations 
\begin{eqnarray}
a=-\frac{16\pi}{F'(B)} \ , \quad  -b+\frac{(a+5b)}{2}=0
\end{eqnarray}
that has solution
\begin{equation}
b=-\frac{a}{3} \  .
\end{equation}
Using these values of $a$ and $b$ we 
obtain inverse relation
\begin{equation}\label{Gammainv}
\Gamma^c_{ba}=-\frac{16\pi}{F'(B)}\bN^c_{ab}+
\frac{16\pi}{3F'(B)}
(\bN^d_{ad}\delta^c_b+\bN^d_{bd}\delta^c_a) \ . 
\end{equation}
Now the Hamiltonian density has the form
\begin{eqnarray}\label{mHF1}
& &\mH^F=\partial_c f^{ab}N^c_{ab}+p^c_B\partial_c B-\mL^F_{quadr}=
(\Gamma_{dc}^d f^{ab}-\Gamma^a_{cd}f^{db}-
\Gamma^b_{dc}f^{da})N_{ab}^c+p^c_B\partial_c B-\mL^F_{quadr}=\nonumber \\
& &=\frac{1}{16\pi}F'(B)[\Gamma^h_{dk}\Gamma^k_{gh}f^{gd}-
\Gamma^f_{fk}\Gamma^k_{gh}f^{gh}]+
\nonumber \\
& &+\frac{1}{16\pi}F''(B)\partial_f B[f^{fa}\Gamma_{ab}^b-\Gamma^f_{ab}f^{ab} ]
-\frac{1}{16\pi}\sqrt{-f}[F(B)-F'(B)B]
\ , \nonumber \\
\end{eqnarray}
using
\begin{eqnarray}
\partial_c f^{ab}=\partial_c \sqrt{-g}g^{ab}+
\sqrt{-g}\partial_c g^{ab}=
\Gamma_{dc}^d f^{ab}-\Gamma^a_{cd}f^{db}-
\Gamma^b_{dc}f^{da} \ , 
\nonumber \\
\end{eqnarray}
where we used the fact that  $\nabla_c \sqrt{-g}=0$ implies
\begin{equation}
\partial_c \sqrt{-g}=\Gamma^d_{dc}\sqrt{-g} \ . 
\end{equation}
In the same way the condition  $\nabla_c g^{ab}=0$ implies
\begin{equation}
\partial_c g^{ab}=-(\Gamma^a_{cd}g^{db}+
\Gamma^b_{cd}g^{da}) \ . 
\end{equation}
Then inserting (\ref{Gammainv}) into (\ref{mHF1}) we obtain 
Hamiltonian density in the form 
%
\begin{eqnarray}\label{mHfbN}
& &\mH_f
=\frac{16\pi}{F'(B)}\bN^h_{dk}f^{gd}\bN^k_{gh}-\frac{16\pi}{3F'(B)}\bN^m_{mk}f^{kg}
\bN^d_{dg}+\nonumber \\
&&+\frac{16\pi}{3F''(B)}p^g_Bf_{gc}
(f^{ab}N_{ab}^c-\frac{F'(B)}{F''(B)}p^c_B)
-\frac{1}{16\pi}\sqrt{-f}[F(B)-F'(B)B] \ .  \nonumber \\
\end{eqnarray}
Finally in order to complete Hamiltonian analysis we have to insert explicit form of $\bN^c_{ab}$ given in
(\ref{bN}) to (\ref{mHfbN}). In fact, after some tedious calculations we find that the Hamiltonian density of $F(R)$ gravity has the form
\begin{eqnarray}
& &\mH^F=
\frac{16\pi}{F'}[N^h_{kd}f^{dg}N^k_{gh}-\frac{1}{3}N^m_{mk}f^{kg}N^n_{ng}]+
\nonumber \\
& &-.\frac{16\pi}{6}F'
[\frac{1}{F''}p^a+\frac{1}{F'}N^a_{mn}f^{mn}]f_{ab}
[\frac{1}{F''}p^b+\frac{1}{F'}f^{rs}N_{rs}^b]
-\frac{1}{16\pi}\sqrt{-f}[F(B)-F'(B)B]
 \ . \nonumber \\
\end{eqnarray}
This is the final form of the covariant Hamiltonian density for $F(R)-$gravity. In order to find corresponding equations of motion we begin with the canonical form of the action
\begin{equation}
S=\int d^4x (N^c_{ab}\partial_cf^{ab}+p^c_B\partial_cB-\mH^F)
\end{equation}
so that its variation has the form 
\begin{eqnarray}
& &\delta S=\int d^4x \left(\delta N^c_{ab}\partial_c f^{ab}+N^c_{ab}\partial_c \delta f^{ab}+
\delta p^c_B\partial_c B+p^c_B\partial_c \delta B \right.-\nonumber \\
&  & \left.-\frac{\delta \mH^F}{\delta N^c_{ab}}\delta N^c_{ab}-
\frac{\delta \mH^F}{\delta f^{ab}}\delta f^{ab}-
\frac{\delta \mH^F}{\delta p^c_B}\delta p^c_B-\frac{\delta \mH^F}{\delta B}\delta B\right)=0
\nonumber \\
\end{eqnarray}
that implies following equations of motion 
\begin{eqnarray}
& &\partial_c f^{ab}=\frac{\delta \mH^F}{\delta N^c_{ab}}=
\frac{16\pi}{	F'}\left[f^{bg}N^a_{gc}+f^{ag}N^b_{gc}\right]
-\frac{16\pi}{3F'}\left[f^{bg}N^n_{ng}\delta^c_a+f^{ag}N^n_{ng}\delta^b_c\right]-
\nonumber \\
& &-\frac{32\pi}{3}f^{ab}f_{ch}\left[\sqrt{-f}\frac{1}{F''}p^h+\frac{1}{F'}
N^h_{rs}f^{rs}\right] \ , \nonumber \\
& &\partial_c N^c_{ab}=-\frac{\delta \mH^F}
{\delta f^{ab}}=-\frac{8\pi}{F'}\left[N^h_{ka}N^k_{bh}+
N^h_{kb}N^h_{ad}\right]+\frac{16\pi}{3F'}N^m_{ma}N^n_{nb}+\nonumber \\
& &-\frac{8\pi F'}{3}
\left[\frac{1}{F''}p^x+\frac{1}{F'}N^x_{mn}f^{mn}\right]f_{xa}
\left[\frac{1}{F''}p^y+\frac{1}{F'}f^{ys}N_{rs}^b\right]f_{yb}+\nonumber \\
& &+\frac{16\pi}{3}\frac{F'}{F''}N_{ab}^x
f_{xy}
\left[\frac{1}{F''}p^y+\frac{1}{F'}N^y_{mn}f^{mn}\right]
+\frac{1}{32\pi}f_{ab}\sqrt{-f}\left[F(B)-F'(B)B\right] \ ,  \nonumber \\
& &\partial_c B=\frac{\delta \mH^F}{\delta p^c_B}=
-\frac{8\pi}{3}\frac{F'}{F''}f_{ab}
\left[\frac{1}{F''}p^b+\frac{1}{F'}N^b_{mn}f^{mn}\right]  \ , 
\nonumber \\
& &\partial_c p^c_B=-\frac{\delta \mH^F}{\delta B}=
\frac{16\pi F''}{F'^2}\left[N^h_{kd}f^{dg}N^k_{gh}-\frac{1}{3}N^m_{mk}f^{kg}N^n_{ng}\right]+
\nonumber \\
& &+\frac{8\pi}{3}F''
\left[\frac{1}{F''}p^a+\frac{1}{F'}N^a_{mn}f^{mn}\right]f_{ab}
\left[\frac{1}{F''}p^b+\frac{1}{F'}f^{rs}N_{rs}^b\right]
+\nonumber \\
& &
-\frac{16\pi}{3}\left[\frac{F'''}{F''^2}p^a+\frac{F''}{F'^2N^a_{mn}f^{mn}}\right]
f_{ab}\left[\frac{1}{F''}p^b+\frac{1}{F'}N^b_{rs}f^{rs}\right]
-\frac{1}{16\pi}\sqrt{-f}F''(B)B \ . \nonumber \\
\end{eqnarray}
Further, the boundary term for $F(R)-$gravity has the form
\begin{equation}\label{mLsurf}
\mL_{surf}^F=-\partial_c\left[\frac{F'}{F''}p^c_B\right]
\end{equation}
which shows an importance of the field $B$ and corresponding conjugate momenta $p_B^c$. On the other hand if we presume that the solution of the equation of motion for $B$ is $B=\mathrm{const}$ we find that  $p^b_B$ is equal $-\frac{F''}{F'}N^b_{mn}f^{mn}$. Inserting this result into (\ref{mLsurf}) we find that it is equal to 
	\begin{equation}
	\mL^F_{surf}=\partial_c[N^c_{mn}f^{mn}	] \  
\end{equation}	
which is the same as in case of pure gravity. Certainly it would be nice to study consequence of this result for thermodynamics aspects of $F(R)-$gravity.
\\
{\bf Acknowledgement:}
\\
The work of JK
is supported by the grant “Integrable Deformations”
(GA20-04800S) from the Czech Science Foundation
(GACR).

\end{document}